\documentclass[12pt]{article} 

\usepackage{graphicx}
\usepackage{amsmath}
\usepackage{amssymb}
\usepackage{hyperref}
\usepackage[height=8.8in,width=6.45in]{geometry}
\usepackage{tikz}
\usepackage{cite}
\usepackage{amscd}
\usepackage{setspace}
\usepackage{bbm}
\usepackage{dsfont}
\usepackage{cancel}
\usepackage{relsize}
\usepackage[utf8]{inputenc}
\usepackage{xcolor}
\newenvironment{Comment}[2]{\noindent\color{#1}{\texttt #2:}}{\par\noindent}

\numberwithin{equation}{section}

\begin{document}
\begin{titlepage}

\begin{flushright}

\end{flushright}

\vskip 3cm

\begin{center}
{\Large \bf
 Vacuum Structures Revisited
}

\vskip 2.0cm

Wei Gu

\bigskip
\bigskip

\begin{tabular}{cc}
Department of Physics, Virginia Tech, 850 West Campus Dr.\\
  Blacksburg, VA 24061, USA
\end{tabular}

\vskip 1cm

{\tt weig8@vt.edu  }

\vskip 1cm

\textbf{Abstract}
\end{center}

We consider the relationship between the higher symmetry and the dynamical decomposition in supersymmetric gauge theory in various dimensions by studying the semi-classical potential energy. We observe that besides the scalar moduli we shall also include the field strength $F_{0\cdots d}$ in the vacuum moduli in the 1+d dimensional theory along with a $\mathbb{Z}_{p}$ $d$-form symmetry. In gauge theory for charge-$p$ matters with this symmetry, we find that the vacua decompose into $p$ different universes at an intermediate scale, which means no dynamical domain wall can interpolate between them. In our setup, we re-derive the existing results on the decomposition in various dimensions. In four dimensions, we propose a UV gauge theory for the generalized super Yang-Mills theory, whose instanton sectors are restricted to the topological number with integer multiples of $p$.

\medskip
\noindent

\bigskip
\vfill

\end{titlepage}

\section{Introduction}\label{Int}
This note discusses the vacuum structure of supersymmetric gauge theory in various dimensions, and we only focus on the theory with a Lagrangian description.\\

Our analysis was motivated by the consideration of the relationship between the higher symmetry \cite{Gaiotto:2014kfa} and the emergent dynamical decomposition in field theory, which was first observed in two-dimensions\cite{Pantev:2005zs, Pantev:2005rh, Pantev:2005wj, Hellerman:2006zs}, and then in other dimensions \cite{ Gu:2021yek, Tanizaki:2019rbk}. Certainly, they can be defined independently. The definition of a $q$-form symmetry in $1+d$ dimensions can be made abstractly, which is implemented by an operator associated with a codimension $q + 1$ closed manifold $M^{^{d-q}}$. The conserved current is a $q+1$-form, and in this note, we are primarily interested in the $d$-form symmetry. One of the fundamental new insights of \cite{Gaiotto:2014kfa} is the explanation of the Higgs mechanism in those ``generalized global symmetries." While the decomposition can be made even in quantum mechanics, for example, consider the double-well potential energy,
\begin{equation}\label{}\nonumber
  V(x)=\lambda\left(x^{2}-a^{2}\right)^{2}.
\end{equation}
There are two interesting limits we can think about. The first one is to take $a$  goes to infinity while $\lambda$ is a finite value, one can, then, compute the fluctuation from the configuration at $x=-a$ to $x=a$ is zero because the energy barrier is infinity. The other limit is to let $\lambda$ goes to infinity while $a$ is a non-vanishing small value. One can easily observe that the wave function must be localized at $x=-a$ or $x=a$. Furthermore, the fluctuation between them is zero because of the same reason.\\

 In quantum field theory, we can have a similar story, if we replace $x$ and $V(x)$ by a scalar field $\phi$ and the semi-classical potential energy density function $U(\phi)$ respectively. It is well-known that parameters in quantum field theory depend on the physical scale. The first limit in the quantum mechanics example has a cousin in the quantum field theory called the ``cluster decomposition." It is a far-infrared behavior. The second one is much more interesting in quantum field theory, if it started at an intermediate physical scale. We denote it as the ``dynamical decomposition." This decomposition is not only physically nontrivial \footnote{In \cite{Tanizaki:2019rbk, Komargodski:2020mxz}, they used the terminology ``universe" in describing this structure to distinguish it from the usual vacua.} but also could lead to interesting mathematical statements reviewed in section \ref{MA}.\\

 As mentioned, the higher symmetry, in general, does not have to be related to the dynamical decomposition directly \footnote{If the quantum theory has symmetry, one can use it to label the Hilbert space and observables by this symmetry. However, this is a trivial ``decomposition" of the vector space, which is not our interest in this note.}. However, there is a special one called $d$-form symmetry that enters our story. In previous studies of the vacuum configuration from the semi-classical potential, people only used scalar fields to parameterize the vacuum because of the Lorentz symmetry in quantum field theory. Now, if we have the $d$-form gauge field, we could also have an extra Lorentz invariant field to label the vacuum configuration. This note includes this in a region where the semi-classical analysis can be trusted. More specifically, we consider the semi-classical potential energy as a function in terms of both scalar fields and field strength $F_{0\cdots d}$ of the $d$-form gauge field,

 \begin{equation}\label{}\nonumber
U\left(\phi, F_{0\cdots d}\right).
\end{equation}
The expectation value of $F_{0\cdots d}$ could teach us the nontrivial information of vacuum configuration, even though it is vanishing \footnote{In contrast, the vanishing of non-scalars is simply because of the Lorentz symmetry.}. In this context, we will find the dynamical decomposition phenomenon in the effective theory with a $\mathbb{Z}_{p}$ $d$-form symmetry.\\

We begin in section \ref{LSMR} by re-discussing the known observations in two dimensions in our context. We try to clarify two points. The first one is where the dynamical decomposition can happen, and the second is why it is different from the ``cluster decomposition" associated with the super-selection rule. Our motivation is to focus on the explanation rather than to produce new results in this section. It can be regarded as a warmup to the more interesting four-dimensional quantum field theory with a three-form $\mathbb{Z}_{p}$ symmetry.\\

In section \ref{CSM}, we discuss the three-dimensional $\cal N$=2 Chern-Simons matter theories. We first re-analyze the dynamical decomposition in the KK-reduction from the three-dimensional gauge theory to the two-dimensional one discussed in \cite{Gu:2021yek} via semi-classical potential energy. Then we comment on why there was no decomposition in the three-dimensional gauge theory. However, we do not claim we have physical proof.\\

In section \ref{GSYM}, we briefly review the four-dimensional generalized super Yang-Mills theory. Then we propose a UV-fundamental theory for  the generalized SYM theory with a $\mathbb{Z}_{p}$ $3$-form symmetry. Finally, we observe the dynamical decomposition from the semi-classical approximation of the UV theory at an intermediate scale. It is similar to the two dimensions.\\

Our last section is devoted to conclusions and future directions.

\section{Linear Sigma Models Revisited}\label{LSMR}
This section investigates the dynamical decomposition in two-dimensional quantum field theory. We do not claim that our results are new, rather, we focus on the physical explanation of decomposition in our setup. The observations found in \cite{Pantev:2005zs, Pantev:2005rh, Pantev:2005wj, Hellerman:2006zs} relied on the computation of massless spectrums and other RG-protected quantities, see also \cite{Sharpe:2014tca, Sharpe:2015mja, Sharpe:2021srf} for more. The interested reader can refer to them for the calculation. Our approach is different, and it supports most observations found in them.\\

\subsection{Charge-$p$ $\mathbb{C}P^{N-1}$ Model}
The ${\cal N}=(2,2) $ linear sigma model for charge-$p$ $\mathbb{C}P^{N-1}$ is a $U(1)$ gauge theory with $N$ gauge charge $p$ chiral superfields, see \cite{Pantev:2005zs, Witten:1993yc} for more details. As mentioned in the introduction, if we have a nontrivial vacuum configuration of $F_{01}$, besides the scalar fields considered in the previous studies, the flowing relevant terms are not necessary vanishing in a vacuum, and they are
\begin{equation}\label{GTL}
{\cal L}=\frac{1}{2e^{2}}F^{2}_{01}+\left(\theta+2n\pi\right) F_{01}-\sum^{N}_{i=1}\mid\phi_{i}\mid^{2}\left(\partial_{\mu}\varphi+p A_{\mu}\right)^{2},
\end{equation}
where we have written each $\phi_{i}$ in the polar variables $\left(\mid\phi_{i}\mid, \varphi_{i}\right)$ defined by $\phi_{i}=\mid\phi_{i}\mid e^{i\varphi_{i}}$. We keep the term $2n\pi F_{01}$ in our Lagrangian because of the Dirac quantization of the abelian group $U(1)$, so $n$ is an integer. Focus on
\begin{equation}\label{GCL}
  {\cal L}_{\varphi}=-\sum_{i}\mid\phi_{i}\mid^{2}\left(\partial_{\mu}\varphi_{i}+p A_{\mu}\right)^{2},
\end{equation}
unfortunately, the factor $\partial_{\mu}\varphi_{i}+p A_{\mu}$ in equation (\ref{GTL}) does not simply depend on the scalar $F_{01}$, although it is gauge-invariant. To see how it relates to the $F_{01}$, we first introduce the auxiliary one-form variables $\lambda_{i,\mu}$, and then we have
\begin{equation}\label{DGCL}
  -\sum _{i}\frac{1}{4\mid\phi_{i}\mid^{2}}\left(\lambda_{i,\mu}\right)^{2}+\epsilon^{\mu\nu}\lambda_{i,\mu}\left(\partial_{\nu}\varphi_{i}+p A_{\nu}\right).
\end{equation}
It is easy to see that integrating out $\lambda_{i,\mu}$ in the above formula will give us the Lagrangian ${\cal L}_{\varphi}$. Now, instead, we integrate out $\varphi_{i}$ first which gives the constraint $\lambda_{i,\mu}=\partial_{\mu}\vartheta_{i}$, where each $\vartheta_{i}$ is a 2$\pi$ periodic variable. Plugging this back into the formula (\ref{DGCL}), we obtain the new Lagrangian\footnote{We assume all of the $\mid\phi_{i}\mid^{2}$ are nonzero, and if some of them vanished, we only take a dual expression for the remaining variables.}
\begin{equation}\label{}\nonumber
  {\cal L}_{\vartheta}=-\sum_{i}\frac{1}{4\mid\phi_{i}\mid^{2}}\left(\partial_{\mu}\vartheta_{i}\right)^{2}-p\vartheta_{i}F_{01},
\end{equation}
where the dual variable $\vartheta$ is coupled to the gauge field $A_{\mu}$ as a dynamical theta angle. The procedure we performed is called the abelian duality \cite{Hori:2000kt}. So we replace the Lagrangian in equation (\ref{GTL}) by a new one
\begin{equation}\label{GTL2}
\widetilde{{\cal L}}=\frac{1}{2e^{2}}F^{2}_{01}+\left(\theta+2n\pi-p\sum^{N}_{i=1}\vartheta_{i}\right)F_{01}-\sum_{i}\frac{1}{4\mid\phi_{i}\mid^{2}}\left(\partial_{\mu}\vartheta_{i}\right)^{2}.
\end{equation}
The Lorentz symmetry does not require the first two terms to vanish in a vacuum. Furthermore, the vacuum configuration, of course, does not depend on the field variables we use. The 2$\pi$ periodicity of each $\vartheta_{i}$ is beautiful connected to the pair creation of heavy particles in the two-dimensional gauge theory \cite{Coleman:1976uz} that changes $\theta$ by 2$lp\pi$ for some integer $l$, so we can restrict $n\in[0, p-1]$. Now, we can write out the semi-classical potential energy
\begin{equation}\label{CPPU}
  U(\phi_{i},F_{01})=\frac{e^{2}_{{\rm eff}}}{2}\left(\sum^{N}_{i=1}p\mid\phi_{i}\mid^{2}-r\right)^{2}+\sum^{N}_{i=1}p^{2}\mid\phi_{i}\mid^{2}\mid\sigma\mid^{2}+\frac{1}{2e^{2}_{{\rm eff}}}F^{2}_{01},
\end{equation}
where we read off the potential energy density in the Hamiltonian formulation. The complex scalar $\sigma$ is the lowest component of the super-field-strength\footnote{The complex scalars $\sigma$ and $\bar{\sigma} $ can be understood from the KK-reduction of the four-dimensional gauge field in $4d$ ${\cal N}=1$ gauge theory, see \cite[sec.~2]{Witten:1993yc} for more details.}, and $r$ is the FI parameter at the physical scale. The last term in equation (\ref{CPPU}) was usually omitted in the previous papers because the gauge field is not a scalar field. In order to find the ground state, i.e. the vanishing configuration of potential energy, the quadratic terms in the potential energy must be vanishing separately because $\mid\phi_{i}\mid^{2}$ and $F_{01}$ are all gauge-invariant Hermitian operators. This suggests that
\begin{equation}\label{GFC}
  \langle F_{01}\rangle=0.
\end{equation}
It looks like we were doing something trivial. However, this is too quick. Let us first look at what is $F_{01}$ in the vacuum. From equation (\ref{GTL2}), we can observe that
\begin{equation}\label{}\nonumber
  F_{01}=e^{2}\left(p\sum^{N}_{i=1}\vartheta_{i}-\theta-2n\pi\right).
\end{equation}
 Then the semi-classical potential energy can be rewritten as
 \begin{equation}\label{CPPU2}
  U(\phi_{i}, \vartheta_{i})=\frac{e^{2}_{{\rm eff}}}{2}\left(\sum^{N}_{i=1}p\mid\phi_{i}\mid^{2}-r\right)^{2}+\sum^{N}_{i=1}p^{2}\mid\phi_{i}\mid^{2}\mid\sigma\mid^{2}+\frac{e^{2}_{{\rm eff}}}{2}\left(p\sum^{N}_{i=1}\vartheta_{i}-\theta-2n\pi\right)^{2},
\end{equation}
 The parameters $e^{2}$ and $r$ are both running under the RG-flow. Although we do not know the exact expression of $e^{2}_{{\rm eff}}$, it becomes the strong coupling in the low energy physics because its mass dimension is two. While the RG-flow of parameter $r$ can be computed exactly which is
\begin{equation}\label{RGFR}
  r=pN\log\frac{\mu}{\Lambda},
\end{equation}
where $\mu$ is a physical scale and $\Lambda$ is the dynamical scale. In order to make sense of the perturbative low energy effective theory NLSM, we restrict our physical scale
\begin{equation}\label{INRGS}
  \Lambda\ll\mu\ll e_{\rm eff}\sqrt{r}.
\end{equation}
We call this the intermediate scale in our note. At this scale, the parameter $r$ is very large suggesting the couplings in NLSM are very small such that the perturbative computation can be trusted. Furthermore, the procedure in integrating out massive modes in the linear sigma model around the vacuum configuration is also reasonable because their masses are proportional to $e_{\rm eff}\sqrt{r}$ which are heavy objects compared to the physical scale. Finally, from equation (\ref{INRGS}), we can find that $e^{2}_{\rm eff}$ approaches infinity at the intermediate scale. This would introduce some nontrivial physics defined in the introduction section called the dynamical decomposition. Let us investigate this by studying the vacuum configuration which can be read off from equation (\ref{CPPU2})
\begin{equation}\label{VCNL}
  \left\langle\sum^{N}_{i=1}\mid\phi_{i}\mid^{2}\right\rangle=\frac{r}{p},\quad\quad \left\langle e^{i\sum^{N}_{i=1}\vartheta_{i}}\right\rangle=e^{i\frac{\theta}{p}+i\frac{2\pi n}{p}},
\end{equation}
for $n=1,2,\ldots p-1$. We have used the fact that all of the variables $\vartheta_{i}$ commute to each other, and only a single-valued field can be the physical observable. We see that different $n$ label different $universes$, and the energy barrier between them is infinity because the parameter $e^{2}_{{\rm eff}}$ goes to infinity. Each universe is a $\mathbb{C}P^{N-1}$, and there is no dynamics that connects one universe to others. Now consider the observable
\begin{equation}\label{}\nonumber
  e^{il*\sum^{N}_{i=1}\vartheta_{i}},
\end{equation}
with an integer $l$. If $l\neq0$ mod $p$, then it is charged by a finite $\mathds{Z}_{p}$ one-form symmetry with the charge $l$. The operation of this symmetry is to shift $\sum^{N}_{i=1}\vartheta_{i}$ by a 2$\pi$. We also call these observables as external probes that connect different universes. If $l=0$ mod $p$, then it is neutral under the one-form symmetry and only acts on a particular universe. The vacua in each universe can be connected by dynamical domain walls or called solitons associated with the superselection rule in a local quantum field theory. \\

We would like to make several comments:
\begin{itemize}
\item The vacuum configuration obtained in equation (\ref{VCNL}) is not in the far-infrared. Therefore, we do not expect any massless Goldstone particle appeared in the usual Higgs mechanism\footnote{In fact, there is no Goldstone particle even in the infinity volume limit in two-dimensional quantum field theory \cite{Coleman:1973ci, Mermin:1966fe}.}. However, the expectation value of $\sum^{N}_{i=1}p\mid\phi_{i}\mid^{2}$ tells us that two bosons, $\sigma$, and $A_{\mu}$, in the vector multiples are massive. This is because there are terms such as $e^{2}\mid\phi\mid^{2}\mid\sigma\mid^{2}$ and $e^{2}\mid\phi\mid^{2}A_{\mu}A^{\mu}$ in the Lagrangian, so their masses are proportional to $e_{{\rm eff}}\sqrt{r}$. A similar argument can apply to gauginos.

\item The non-vanishing expectation value of $e^{i\sum^{N}_{i=1}\vartheta_{i}}$ does not suggest that the flavor symmetry is broken. We still have $N-1$ free field variables which are generators of the group $U(1)^{N-1}$, and this group is the max-torus of the flavor symmetry of the target.

\item We can also consider the linear sigma model with a superpotential, and the dynamical decomposition would not be changed. Because this phenomenon is due to the existence of the $\mathbb{Z}_{p}$ one-form symmetry and the strong coupling limit of the gauge coupling $e^{2}_{{\rm eff}}$. Furthermore, the superpotential only introduces the usual F-term which, of course, does not affect the story that happened in the D-term.

 \item One may also worry that the full nonperturbative quantum correction would affect our statement on dynamical decomposition found in the semi-classical potential energy. However, one can easily estimate that the contribution of the degree one instanton is proportional to $q=e^{-r+i\theta}$, which is close to zero, and the higher degrees instantons are higher-order power of $q$. So one can expect that the summation of nonperturbative correction gives a finite convergence series, and this is what Hori and Vafa \cite{Hori:2000kt} originally performed in deriving the abelian mirror symmetry. An expert would already notice that our dual variables $\vartheta_{i}$ are the usual field variables in the mirror. A beautiful story of mirror symmetry is that the potential energy in the mirror is exact, so it is better to discuss the decomposition in the mirror. See section \ref{MDNCP}.

\item  Our last comment in this section is whether we have a dynamical decomposition in the linear sigma model. Our answer would be $no$. This follows our definition of dynamical decomposition because, in the UV, the gauge coupling is a small parameter. One can still have an extra label on the physical observables and state spaces because of the one-form symmetry, and we call them sectors. However, this is a trivial vector space decomposition in a quantum theory\footnote{In quantum mechanics, we usually find a complete set of commuting operators to label the Hilbert space.}. Furthermore, the physical operators in the gauge theory do not depend on the gauge coupling, which means we can compute the NLSM correlation functions in the linear sigma model. The decomposition of correlation functions only follows the decomposition of the nonlinear sigma model that does not imply we have a decomposition in the linear sigma model. Finally, we want to mention two different field configurations which connect sectors. The first one, in the UV, is the non-BPS domain wall configuration that fluctuates from one vacuum in one sector to another vacuum in a different sector\footnote{The origin of scalar field spaces is the bridge in connecting different universes because there is no order parameter here to label the different ``universe."}. The tension is approximately proportional to $2\pi^{2}e^{2}_{{\rm eff}}$, and one can notice that they are infinity heavy in the infrared. The second field configuration that appeared in the UV gauge theory is the $point$-$like$ instanton \cite{Witten:1993yc}, and one can show that they can not be decomposed into different sectors. However, because they are positive codimension objects \cite{Morrison:1994fr}, they do not affect the correlation functions of NLSM on charge-$p$ projective space. Furthermore, the masses of point-like instantons are $\sim$ $e_{\rm eff}\sqrt{r}$, which means they are also infinity massive in the low energy physics. All these objects suggest that we do not have a decomposition in the extreme UV, but we have a perfect decomposition at the intermediate scale.

\end{itemize}

\subsection{Mirror of NLSM on Charge-$p$ $\mathbb{C}P^{N-1}$ Model}\label{MDNCP}

As commented in the previous section, it is better to discuss the dynamical decomposition in the mirror. Following \cite{Hori:2000kt}, we first define the mirror Landau-Ginzburg model: the target space $\left(\mathbb{C}^{\ast}\right)^{N}\times\mathbb{C}$ with the superpotential
\begin{equation}\label{}\nonumber
  W=\Sigma\left(p\sum^{N}_{i=1} Y_{i}-t\right)+\sum^{N}_{i=1}e^{-Y_{i}}.
\end{equation}
The K\"{a}hler potential is
\begin{equation}\label{}\nonumber
  K\left(Y_{i},\Sigma;\bar{Y_{i}},\bar{\Sigma}\right)=-\frac{1}{2e^{2}}\bar{\Sigma}\Sigma-\sum^{N}_{i=1}\frac{1}{2}\left(Y_{i}+\bar{Y_{i}}\right)\log\left(Y_{i}+\bar{Y_{i}}\right).
\end{equation}
Writing the Lagrangian in terms of the component fields of chiral super fields $Y$ and $\Sigma$, there is a term like
\begin{equation}\label{}\nonumber
  p\vartheta_{i}F_{01},
\end{equation}
in the Lagrangian.\footnote{In the mirror, the component field $F_{01}$ in the chiral superfield $\Sigma$ is not a gauge field strength, although we use the same notation.} Because of the 2$\pi$ periodicity of $\vartheta_{i}$, we have the Dirac quantization condition
\begin{equation}\label{}\nonumber
  \frac{p}{2\pi}\int F\in \mathbb{Z}.
\end{equation}
Now, we can write out the exact potential energy
\begin{equation}\label{}\nonumber
  U(y_{i}, \sigma)=\frac{e^{2}}{2}\left| p\sum^{N}_{i=1}y_{i}-t+2i\pi n\right|^{2}+\sum^{N}_{i=1}\left| p\sigma-e^{-y_{i}}\right|^{2},
\end{equation}
where $\sigma$ is the lowest component of the chiral superfield $\Sigma$, while $y_{i}$ is the lowest component of the chiral superfield $Y_{i}$,
\begin{equation}\label{}\nonumber
  y_{i}=\varrho_{i}-i\vartheta_{i}.
\end{equation}
One may already notice that $\varrho_{i}$ is equal to the dual variable $\mid\phi_{i}\mid^{2}$ in the linear sigma model. Then, from the exact potential, we can observe the same dynamical decomposition found in the gauge theory. Although we only have F-terms in the mirror, however, the K\"{a}hler potential of $\Sigma$-field provides the coefficient $e^{2}$ in the potential energy. The full nonperturbative correction only changes the expectation value of $\sigma$ field and does not affect the decomposition conclusion.  See also \cite{Gu:2018fpm, Chen:2018wep, Gu:2020ivl} for the dynamical decomposition in the nonabelian mirrors and \cite{Seiberg:2010qd, Komargodski:2017dmc, Anber:2018jdf, Armoni:2018bga, Misumi:2019dwq, Tanizaki:2019rbk, Cherman:2020cvw,Komargodski:2020mxz } for the dynamical decomposition in the charge-$p$ Schwinger model. Finally, we want to comment that if $e^{2}$ is very large, the $\Sigma$-field's K\"{a}hler potential is suppressed, which can be treated as an auxiliary field. Then we can integrate out it to introduce delta functions in the $Y$-field space, there are $p$ different sectors. They can not talk to each other, because they localize at different delta functions. In the next section, we will give a brief review of the application of decomposition to math.

\subsection{Mathematical Applications}\label{MA}
In the previous two sections, we have reviewed a finer vacuum structure called the dynamical decomposition. Certainly, it is a nontrivial physical statement, which relates to several quantum field theoretic aspects such as higher symmetry, 't Hooft anomaly, etc. Furthermore, it can be applied to the string compactification \cite{Hellerman:2006zs}. However, in this section, we mainly focus on its applications to math.\\

The dynamical decomposition is a statement of the physical two-dimensional nonlinear sigma model. So the Witten-type topological quantum field theories, A-twisted or B-twisted of the physical NLSM, shall share the same conclusion. On the one hand, the topological A-model in math is related to the Gromov-Witten theory and Fukaya category. The dynamical decomposition nontrivially suggests that Gromov-Witten invariants \cite{Andreini:2008AJT, Andreini:2009AJT} and Fukaya categories have decomposition properties if the target space manifold has a $\mathbb{Z}_{p}$ one-form symmetry. The Gromov-Witten theory is concerned with intersection numbers on the moduli space of stable map (from the genus-zero curve to the variety), and these invariants can be encoded into a $Q$ series
\begin{equation}\label{}
  \sum_{d=0}n_{d}Q^{d},
\end{equation}
where the rational number $n_{d}$ is related to the Gromov-Witten invariants. The decomposition, in this context, means the moduli space of stable map can be decomposed into the stable map from the genus-zero curve to $p$ same target manifold but with $p$ different Novikov variables, and they are $Q^{\frac{1}{p}}e^{\frac{i2n\pi}{p}}$, for $n\in[0,\ldots,p-1]$. Therefore, only the map with a degree $pd$ ($d$ is a non-negative integer) can contribute to the invariants. While the categories, in mathematics, consists of a collection of ``objects" that are linked by ``morphisms." The decomposition of a category means that objects in this category can be decomposed into $p$ sets, and the ``morphisms" only link the objects in the same set. On the other hand, the topological B-model in math is related to the Picard-Fuchs equation and derived category of coherent sheaves. The decomposition of a Picard-Fuchs system says the solution space of this PF-equation with an order $pn$ can be decomposed into $p$ sectors, and each sector consists of the $n$-dimensional solution space of an order $n$ Picard-Fuchs equation. Some studies of derived categories of coherent sheaves of weighted projective space can be found in the math literature \cite{Auroux:2004AKO}. However, a complete survey of the decomposition in categories is still worth pursuing in math.\\

We close this section by mentioning that we can also construct the A-twisted/B-twisted linear sigma model. The observables in the A-model do not depend on $e^{2}$, so the Gromov-Witten invariant can be computed in the A-twisted linear sigma model, which could have a decomposition phenomenon. There is no definition of Fukaya categories of linear sigma model in the literature yet, however, one may expect they do not depend on the parameter $e^{2}$, so they could have a dynamical decomposition if the theory has a nontrivial one-form symmetry. The topological B-model of linear sigma model is also interesting. First, we still do not have a definition of B-twisted closed string linear sigma model yet, however, the B-twisted boundary in the linear sigma model has been studied in \cite{Govindarajan:2000ef, Hori:2000ic, Hellerman:2001ct, Herbst:2008jq, Hori:2013ika} and its math theory, the derived category of a GIT quotient and its quotient stack\footnote{We thank Ming Zhang for pointing out the correct terminology we should use here.}, has been built in \cite{Leistner:2012ika}. Furthermore, one lesson from these studies is that the derived category is not RG-protected, see also \cite{Kapustin:2002bi}. Thus from the physical reason, we should not expect the derived category of gauge theory has the decomposition feature \footnote{Consider, for example, the linear sigma model for a projective space. The derived category of this GLSM, (in math terminology it is the derived category of stack quotient $[\mathbb{C}^{N}/U(1)]$), consists of, besides the data from the projective space, an extra part from the unstable locus where all of the coordinates are vanishing.}, although its stable part can be decomposed because it corresponds to B-branes dynamics of the nonlinear sigma model. We leave this proof to the interested mathematician.

\section{3d ${\cal N}=2$ Chern-Simons Matter Theories}\label{CSM}
In this section, we discuss the vacuum structure in the three-dimensional gauge theory, and we mainly focus on the ${\cal N}=2$ Chern-Simons matter theories. However, we expect that there should be a similar story in other 3d gauge theories.\\

As discussed in the introduction, the vacuum configuration is, not only, labeled by the scalar moduli, but also the field strength due to the two-form symmetry. So in 3d ${\cal N}=2$ gauge theory for charge-$p$ $\mathbb{C}P^{N-1}$, we do not expect the $\mathds{Z}_{p}$ one-form symmetry causes any nontrivial dynamical decomposition of vacua. However, as first discussed in \cite{Gu:2021yek}, the 3d gauge theory for charge-$p$ $\mathbb{C}P^{N-1}$ will decompose into $p$ 2d linear sigma models for the same target via the study of the KK-reduction of the exact twisted superpotential. In this note, we review this fact by using the semi-classical potential energy. \\

Following \cite{Aganagic:2001uw} we relate the 3d data to the 2d's by KK-reduction. We first compactified the spacetime on ${\rm R}^{2}\times S^{1}$, and the circle $S^{1}$ has the 2$\pi$R periodic. Then the 2d matter field can be defined from the 3d one's as
\begin{equation}\label{}\nonumber
  \Phi_{2d}=\sqrt{2\pi R}\Phi_{3d},
\end{equation}
with twisted masses given by $\frac{in}{R}$, for integer $n$. While the complex scalar field $\sigma=\sigma_{1}+i\sigma_{2}$ in the two-dimensional vector multiplet can be descended from the scalar $\widetilde{\phi}$ and the circle component of gauge field $v^{\mu}$ in the three-dimensional abelian vector multiplet:
\begin{equation}\label{}\nonumber
  \sigma_{1}=\frac{1}{2\pi R}\int_{S^{1}}\widetilde{\phi},\quad\quad  \sigma_{2}=\frac{1}{2\pi R}\int_{S^{1}}v^{2}\equiv\sigma_{2}+\frac{1}{R},
\end{equation}
where the periodicity of the Wilson line $\sigma_{2}$ arises from large gauge transformations. The 2d bare couplings also can be reduced from 3d's by
\begin{equation}\label{}\nonumber
  \frac{2\pi R}{e^{2}_{3{\rm d}}}=\frac{1}{e^{2}_{2{\rm d}}},\quad\quad r_{2d}=2\pi R r_{3d}.
\end{equation}
Now, we can write out the potential energy of the three-dimensional gauge theory
\begin{eqnarray}\label{P2F3}
  U\left(\phi_{i}, \vartheta_{i}\right)&=&\frac{e^{2}_{3d}}{4\pi R}\left(\sum^{N}_{i=1}p\mid\phi_{i}\mid^{2}-r\right)^{2}+\frac{e^{2}_{3d}}{4\pi R}\left(p\sum^{N}_{i=1}\vartheta_{i}-\theta-2n\pi\right)^{2} \\\nonumber
   &+& \sum^{N}_{i=1}\mid\phi_{i}\mid^{2}\left| p\sigma_{1}+i\left(p\sigma_{2}+\frac{n}{R}\right)\right|^{2},
\end{eqnarray}
where we have suppressed the information of the Chern-Simons level, but we have already chosen a proper one such that there is no topological vacuum in the Higgs phase \cite{Gu:2021yek, Intriligator:2013lca}. From equation $\left(\ref{P2F3}\right)$, we can find the vacuum configuration
\begin{equation}\label{}\nonumber
  \left\langle \sum^{N}_{i=1}p\mid\phi_{i}\mid^{2} \right\rangle=r,\quad\langle\sigma_{1}\rangle=0,\quad \left\langle e^{2i\pi R\sigma_{2}}\right\rangle=e^{2i\pi\frac{n}{p}}.
\end{equation}
Besides the charged-$p$ projective space moduli, we have an extra label due to the Wilson loop $e^{2i\pi R\sigma_{2}}$ \cite{Witten:1999ds,Gu:2020zpg}. Furthermore, one can show that
\begin{equation}\label{}\nonumber
  e^{l\ast2i\pi R\sigma_{2}}
\end{equation}
is a representation of the $\mathds{Z}_{p}$ one-form symmetry with the charge $l$ mod $p$ \cite{Gaiotto:2014kfa, Witten:1999ds}. \\

Now, take the vanishing limit of radius $R$ and keep the gauge coupling $e^{2}_{2d}$ small, then we can observe that the 3d gauge theory decomposes into $p$ universes, and each of them is a 2d linear sigma model for charge-$p$ $\mathbb{C}P^{N-1}$. Furthermore, following section \ref{LSMR}, each universe will further decompose into $p$ universes when the gauge coupling becomes large.\\

Now, come back to the question mentioned at the beginning: do we have a dynamical decomposition in the 3d gauge theory? We suspect there is no decomposition in the three-dimensional gauge theory, however, we do not claim we have physical proof. The reason we expect this is we do not have theta angle in three-dimensional gauge theory\footnote{One can certainly turn on the two-form symmetry in the three dimensions. However, even it has the decomposition feature due to this two-form symmetry, we expect that it would not talk to our Chern-Simons matter theories.}. Mathematically, it is because the second homotopy group of any gauge group is trivial.

\section{Four-dimensional ${\cal N}=1$ Generalized Super Yang-Mills Theory}\label{GSYM}
The generalized four-dimensional Yang-Mills theory with $3$-form symmetry was first proposed in \cite{Seiberg:2010qd}, where Seiberg modified the instanton sectors to be a multiple of $p$ by adding to the Lagrangian
\begin{equation}\label{}\nonumber
  i\vartheta\left(\frac{1}{8\pi^{2}}{\rm tr}\left[F\left(a\right)\wedge F\left(a\right)\right]-\frac{p}{2\pi}F^{(4)}\right),
\end{equation}
where $\vartheta$ is a Lagrange multiplier and $F^{(4)}=d A^{(3)}$. This theory has not only the $\mathbb{Z}_{N}$ one-form symmetry but also a $\mathbb{Z}_{p}$ 3-form symmetry. The systematical study of this generalization can be found in \cite{Misumi:2019dwq}. Furthermore, the authors in \cite{Misumi:2019dwq} extended the story to the generalized ${\cal N}=1$ SYM theory. They uncover the dynamical decomposition in this system by computing 't Hooft anomaly incorporated with the intriguing higher-group structure. They observed that the $Np$ vacua split into $p$ universes with $N$ vacua in each. Furthermore, they found there are dynamical domain walls in between two vacua of a universe, but no dynamical domain wall connects two different universes. The remaining question is whether the auxiliary field $\vartheta$ can be descended from a dynamical field. Following the same idea used in two dimensions in section \ref{LSMR}, we propose a 4d UV ${\cal N}=1$ generalized super Yang-Mills theory that all of the fields are dynamical fields.\\

We first write out the known Lagrangian of ${\cal N}=1$ super Yang-Mills theory
\begin{eqnarray}
  S_{{\rm SYM}}&=& \frac{1}{2g^{2}}\int {\rm tr}\left[F(a)\wedge\star F(a)\right]+\frac{i\theta_{{\rm YM}}}{8\pi^{2}}\int {\rm tr}\left[F(a)\wedge F(a)\right] \\\nonumber
  &+&  \overline{\lambda}\overline{\sigma}^{\mu}\left(\partial_{\mu}\lambda+i\left[a_{\mu},\lambda\right]\right).
\end{eqnarray}
The system has a $\mathbb{Z}_{N}$ one-form symmetry. The $\mathbb{Z}_{N}$ two-form gauge field can be realized as a pair of $U(1)$ one-form and two-form gauge fields, $B^{(1)}$, $B^{(2)}$, with the constraint,
\begin{equation}\label{}\nonumber
  NB^{(2)}=dB^{(1)},
\end{equation}
where the one-form symmetric transformation is
\begin{equation}\label{}\nonumber
  B^{(2)}\mapsto B^{(2)}+d\Lambda^{(1)}, B^{(1)}\mapsto B^{(1)}+N\Lambda^{(1)}.
\end{equation}
In order to see this symmetry from the Lagrangian, following \cite{Gaiotto:2014kfa, Kapustin:2014gua}, we introduce the $U(N)$ gauge field, $\widetilde{a}$, and relate it to the dynamical $SU(N)$ gauge field locally as
\begin{equation}\label{}\nonumber
  \widetilde{a}=a+\frac{1}{N}B^{(1)}.
\end{equation}
Then we replace the $SU(N)$ field strength with the gauge-invariant combination of the $U(N)$ field strength $F\left(\widetilde{a}\right)$ and $B^{(2)}$:
\begin{equation}\label{}\nonumber
  F(a)\equiv F\left(\widetilde{a}\right)-B^{(2)}.
\end{equation}
One can show that $F(a)$ is manifestly invariant under the one-form symmetric transformation. However, this system has a mixed anomaly between $\theta$-angle periodicity and $\mathbb{Z}_{N}$ one-form symmetry. More specifically, the partition function is not a single-valued function under the shifting of $\theta_{{\rm YM}}$ by a $2\pi$, i.e,
\begin{equation}\label{}\nonumber
  Z_{{\rm SYM}}\left(\theta_{{\rm YM}}+2\pi\right)=\exp\left(i\frac{N}{4\pi}\int B^{(2)}\wedge B^{(2)}\right)Z_{{\rm SYM}}\left(\theta_{{\rm YM}}\right),
\end{equation}
where we have the non-trivial 't Hooft flux
\begin{equation}\label{THDQ}
  \frac{N}{8\pi^{2}}\int B^{(2)}\wedge B^{(2)}\in\frac{1}{N}\mathbb{Z}.
\end{equation}
It was observed in \cite{Seiberg:2010qd} that in the generalized Yang-Mills theory, the instanton sectors have been modified to
\begin{equation}\label{}\nonumber
  \left(\frac{1}{8\pi^{2}}{\rm tr}\left[F\left(a\right)\wedge F\left(a\right)\right]-\frac{p}{2\pi}F^{(4)}\right)=0,
\end{equation}
where $F^{(4)}$ is a four-form field strength with the normalization condition $\int F^{(4)}\in 2\pi \mathbb{Z}$. This suggests that
\begin{equation}\label{THDQ2}
  \frac{N}{8\pi^{2}}\int B^{(2)}\wedge B^{(2)}\in \mathbb{Z},
\end{equation}
which does not agree with the quantization condition in (\ref{THDQ}). Hence, we can not gauge the $\mathbb{Z}_{N}$ one-form symmetry solely in the generalized super Yang-Mills theory. This puzzle has been resolved in \cite{Misumi:2019dwq} by gauging the $\mathbb{Z}_{p}$ three-form symmetry at the same time, and a general statement for the $d+1$ group structure in $1+d$-dim QFT has been proved in \cite{Tachikawa:2017gyf}. The interested reader can read them for more details. In the remaining section, we propose a UV model that reduces to the generalized super Yang-Mills at the intermediate scale.\\

We first make the three-form gauge field dynamics by introducing a kinetic term
\begin{equation}\label{}
  \frac{1}{2e^{2}_{4d}}F^{(4)}\wedge \star F^{(4)}.
\end{equation}
 The mass dimension of $e^{2}_{4d}$ is 4, so it is a weak coupling in the UV and becomes the strong coupling in the IR. Following the discussion of the one-form symmetry in the four-dimensional Maxwell theory in \cite{Kapustin:2014gua}, one can observe that our generalized Maxwell equation has a three-forms $U(1)$ global symmetry. In order to have the correct degrees of freedom at the low energy physics, we also need to introduce a matter field $\phi$ charged by the two-forms $U(1)$ symmetry with the charge $p$ \footnote{It is consistent with the fact the instantons are classified by degree 4 cohomology.}, such that the symmetry in low energies is a three-forms $\mathbb{Z}_{p}$ global symmetry. The Lagrangian of the charged matter is
\begin{equation}\label{}
  \left|D_{\mu}\phi\right|^{2}.
\end{equation}
Since this ``magnetic" charged matter has not appeared in the previous studies compared to the usual ``electronic" charge matters, we would like to give more details about the gauging process. \\

To start, let us consider a free complex scalar $\phi$ in 1+d dimensions. We can write the field in polar coordinates $\phi=\left|\phi\right| e^{i\varphi}$ except at the origin. Without losing the generality, we restrict to the Lagrangian of the phase factor $\varphi$:
\begin{equation}\label{}
  -d\varphi\wedge \star d\varphi.
\end{equation}
This has a manifest global zero-form symmetry $\varphi\mapsto \varphi+p\alpha$ for a constant $\alpha$. Then perform a local variation, we can find the conserved current of this zero-form symmetry
\begin{equation}\label{}
  -2p\star d\varphi.
\end{equation}
To gauge this zero-form symmetry, we couple to a one-form $U(1)$ gauge field $A$ with the gauge-invariant action
\begin{equation}\label{}
  -\left(d\varphi+p A\right)\wedge \star \left(d\varphi+p A\right).
\end{equation}
In 1+1 dimensions, the above procedure gives the same expression as equation (\ref{GCL}) up to an overall radial variable. \\

The free theory actually has an extra (d-1)-form global symmetry with the conserved current:
 \begin{equation}\label{MCJ}
  (-)^{d}2 p d\varphi.
\end{equation}
It is difficult to see this symmetry manifest in the ``electronic" description; instead, it shifts the dual $(d-1)$-form variable by a $(d-1)$-form flat ``connection": $C^{(d-1)}$. However, we still can gauge this symmetry as usual in the electronic variable. We first couple the current to a $d$-form field $A^{(d)}$:
\begin{equation}\label{GHCP}
  (-)^{d}2p A^{(d)}\wedge d\varphi.
\end{equation}
This action is not yet gauge-invariant for a local gauge transformation. To get the gauge invariant one, we should introduce an extra term $p^{2}A^{(d)}\wedge \star A^{(d)}$ in the Lagrangian. If we put it all together we obtain
\begin{equation}\label{}
    -\left( d\varphi+p  (-)^{d}\star A^{(d)}\right)\wedge \star \left(d\varphi+p (-)^{d}\star A^{(d)}\right).
\end{equation}
This section mainly focuses on the $d=3$ case, which means we are gauging the two-form $U(1)$ global symmetry. Furthermore, we are interested in a theory with the nontrivial instanton sector which means the mass dimension of our gauge field strength $F_{0123}$ is four. Therefore, we expect the mass dimension of the radial part, $\left|\phi\right|$, is vanishing from equation (\ref{GHCP}). This also indicates, from the current (\ref{MCJ}), that the non-vanishing two-form conserved charge is infinity heavy in the far infrared. Thus, it does affect our usual infrared scattering amplitudes. Our gauge theory is, of course, a UV-fundamental theory. However, the matter has an unusual kinetic term
\begin{equation}\label{}\nonumber
  d\left|\phi\right|\wedge \star d\left|\phi\right|,
\end{equation}
which has mass dimension two. Therefore, the coefficient of this kinetic term is not actually a dimensionless constant, rather, it has the mass dimension two. This means the only finite energy fluctuation in the far-infrared is the constant mode. One can compute the one-point correlation function of $\left|\phi\right|^{2}$ at the physical scale $\mu$:
\begin{equation}\label{EPM}
  \left\langle\left|\phi\right|^{2}\right\rangle\cong\int_{\mu\leq|k|\leq\Lambda_{{\rm UV}}} \frac{d^{4}k}{(2\pi)^{4}}\frac{1}{k^{4}}\cong\log\left(\frac{\Lambda_{{\rm UV}}}{\mu}\right).
\end{equation}
The propagator $k^{-4}$ in momentum space is an unusual one. It can be understood by replacing our variable $\left|\phi\right|$ with $\left|\partial^{-1}_{\mu}\widetilde{\phi}\right|$, where the field $\widetilde{\left|\phi\right|}$ has the usual propagator $k^{-2}$. There could be an overall constant off. However, the sign of this constant can be absorbed into the sign of charge $p$, while the absolute value of this constant can be absorbed into the scales.
\\

We can, furthermore, turn on the FI parameter of this higher $U(1)$ symmetry
\begin{equation}\label{}
  -r_{4d}\cdot D,
\end{equation}
where $D$ is an auxiliary field. The $D$-term coupling is
\begin{equation}\label{}
  D\left(p\left|\phi\right|^{2}_{{\rm Bare}}-r^{{\rm UV}}_{4d}\right).
\end{equation}

Because of SUSY, we claim the RG-flow of $r_{4d}$ only receive a one-loop correction:
\begin{equation}\label{}
  r_{4d}\left(\mu\right)=r^{{\rm UV}}_{4d}+p\log\left(\frac{\mu}{\Lambda_{{\rm UV}}}\right).
\end{equation}
This correction can be understood from the calculation in equation (\ref{EPM}). \\

 The last term we introduce is
\begin{equation}\label{}
 \left(\frac{ \widetilde{\theta}}{2\pi}+ n\right)\int F^{(4)}.
\end{equation}
The integer $n$ follows the Dirac quantization of the flux $F^{(4)}$. To affect the instanton sectors of the super Yang-Mills theory, we must impose a condition on theta angles:
\begin{equation}\label{}
  \widetilde{\theta}\equiv p \theta_{{\rm YM}}.
\end{equation}
Therefore, besides the dynamical scale in Yang-Mills theory $\Lambda$, here we have the second RG-invariant dynamical scale defined by
\begin{equation}\label{}
  \widetilde{\Lambda}\equiv \Lambda_{{\rm UV}} \exp\left(-r^{{\rm UV}}_{4d}+i\widetilde{\theta}\right)\gg \Lambda.
\end{equation}
Hence, it is possible that the theory still stays at the perturbative region of Yang-Mills theory even though the coupling $e^{2}_{4d}$ is large. After gauging the one-form $\mathbb{Z}_{N}$ symmetry, the theory shall have nontrivial 't Hooft flux (\ref{THDQ}). In order to make this possible, we require that $A^{(3)}$ transforms under the one-form gauge transformation as \cite{Kapustin:2014gua,Gaiotto:2014kfa, Misumi:2019dwq}
\begin{equation}\label{}
 p A^{(3)}\mapsto p A^{(3)}-i\left(\frac{N}{2\pi}B^{(2)}\wedge \Lambda^{(1)}+\frac{N}{4\pi}\Lambda^{(1)}\wedge d\Lambda^{(1)}\right).
\end{equation}
One can find that $\frac{i}{4\pi}\int {\rm tr} F(\widetilde{a})\wedge F(\widetilde{a})+p\int F^{(4)}$ is gauge-invariant under the one-form symmetry. This means the nontrivial 't Hooft flux is physically reasonable.\\

The vacuum structure, besides the $D$-term, also depends on the following three terms
\begin{equation}\label{}
   \frac{1}{2e^{2}_{4d}}F^{(4)}\wedge \star F^{(4)}+\left(\frac{ \widetilde{\theta}}{2\pi}+ n\right) F^{(4)}+ \left|D_{\mu}\phi\right|^{2}.
\end{equation}
Following section \ref{LSMR}, we use a dual variable to rewrite the formula as
\begin{equation}\label{}
  \frac{1}{2e^{2}_{4d}}F^{(4)}\wedge \star F^{(4)}+\left(\frac{ \widetilde{\theta}-p\vartheta}{2\pi}+ n\right) F^{(4)}-\frac{1}{4\mid\phi\mid^{2}}\star d\vartheta\wedge d\vartheta.
\end{equation}
From the equation of motion, we have
\begin{equation}\label{}
  \star F^{(4)}=e^{2}_{4d}\left(\frac{p\vartheta-\widetilde{\theta}}{2\pi}-n\right).
\end{equation}

Finally, the semi-classical potential energy of our UV gauge theory is
\begin{equation}\label{PGSYM}
  U\left(\star F^{(4)}, \phi\right)=\frac{e^{2}_{4d,{\rm eff}}}{2}\left(p\mid\phi\mid^{2}-r_{4d}\right)^{2}+\frac{2\pi^{2}}{e^{2}_{4d,{\rm eff}}}\left(\star F^{(4)}\right)^{2},
\end{equation}
where $r_{4d}=p\log\frac{\mu}{\widetilde{\Lambda}}$.  Now, consider the intermediate physical scale that
\begin{equation}\label{}
 e_{4d}\sqrt{r_{4d}} \gg \mu\gg \widetilde{\Lambda}\gg\Lambda,
\end{equation}
where our perturbative theory is physics reasonable. The vacuum configuration from (\ref{PGSYM}) is
\begin{equation}\label{}
 \left\langle \mid\phi\mid^{2}  \right\rangle=\frac{r_{4d}}{p},\quad\quad \left\langle e^{i\vartheta}  \right\rangle=e^{i\frac{2n\pi+\widetilde{\theta}}{p}},
\end{equation}
for $n=0,\ldots,p-1$. One can easily find that the theory has $p$ different universes labeled by the integer $n$.
The operator
\begin{equation}\label{}\nonumber
 e^{il*\vartheta}
\end{equation}
is charged under the $\mathbb{Z}^{(3)}_{p}$ symmetry with the charge $l$ mod $p$. The dynamical decomposition can be easily seen from our semi-classical analysis of the UV gauge theory we proposed. However, one drawback is that we could not see all $Np$ vacua explicitly in the semi-classical analysis. For this purpose, we must include the full nonperturbative quantum correction in the potential energy to see the entire vacuum configuration. However, we do not perform this calculation in this note as this would not bring any new insight into the dynamical decomposition. Rather, we see that the radial direction of $\phi$ is heavy that can be integrated out. Furthermore, the kinetic term of $\vartheta$ is suppressed, which means it is an auxiliary field at the intermediate scale. Finally, we can estimate the tension of the non-BPS domain wall between different sectors is proportional to $e^{2}_{4d}$ which suggests that the domain wall becomes infinity heavy even at the intermediate scale. Therefore, our 4d UV gauge theory is correctly reduced to the effective theory proposed in \cite{Seiberg:2010qd} and studied in \cite{Misumi:2019dwq}. The vacua of the UV gauge theory can, then, be described explicitly in terms of the low energy degrees of freedom.\\

Although it is not crucial for our decomposition story; however, we end this section by proposing the full supersymmetric Lagrangian of our UV gauge theory.  We first start with the chiral superfield defined as usual
\begin{equation}\label{}
  \Phi=\phi\left(y\right)+\Theta\sqrt{2}\psi\left(y\right)+\Theta\Theta F(y),
\end{equation}
where $\Theta$ is the Grassmann odd coordinate in the superspace, and $y^{\mu}=x^{\mu}+i\Theta\sigma^{\mu}{\bar \Theta}$. Our notation follows the book by Wess and Bagger \cite{WB92}. It turns out that the most difficult technical part of defining our Lagrangian is the super $p$-form vector multiplet; however, it has been already systematically studied in the context of $4d$ ${\cal N}=1$ theory four decades ago \cite{Gates:1980ay}.  We only focus on the super three-form vector multiplet in this note, let us denote the scalar vector multiplet to be $V$, then following \cite[equation(3.12)]{Gates:1980ay}, we define the super-field strength
\begin{equation}\label{}
  \bar{\Sigma}=\frac{1}{4}D^{2}V={\bar\sigma}+i\sqrt{2}{\bar\Theta\bar\lambda}+{\bar\Theta\bar\Theta}\left(D+iF_{0123}\right),
\end{equation}
where we use a different notation and possibly a different normalization of each component field. Furthermore, the mass dimensions of fields are also different. For example, we require the $\sigma$-field to have the mass dimension three. The kinetic term of gauge field is
\begin{eqnarray}
  {\cal L}_{{\rm vector}} &=& \frac{1}{2e^{2}_{4d}}\int d^{2}\Theta d^{2}{\bar \Theta}\bar{\Sigma}\Sigma \\\nonumber
   &=& \frac{1}{2e^{2}_{4d}}\left(\partial^{\mu}\bar{\sigma}\partial_{\mu}\sigma+i\partial_{\mu}\bar{\lambda}{\bar\sigma}^{\mu}\lambda+D^{2}+F^{(4)}\wedge \star F^{(4)}\right),
\end{eqnarray}
where $\sigma^{\mu}$ is the Pauli matrix.  The kinetic term of the charged matter is also straightforward
\begin{eqnarray}
  {\cal L}_{{\rm matter}} &=& \int d^{2}\Theta d^{2}{\bar \Theta}\bar{\Phi}e^{pV}\Phi \\\nonumber
   &=&|D_{\mu}\phi|^{2}-i{\bar\psi}{\bar\sigma}^{\mu}D_{\mu}\psi\\\nonumber
   &&+Dp|\phi|^{2}+|p\sigma|^{2}|\phi|^{2}+|F|^2+p{\bar\sigma}{\bar F}\phi+pF\sigma\bar{\phi}+ip{\bar \psi}{\bar\lambda}\phi+ip{\bar \phi}\lambda\psi.
  \end{eqnarray}
The last term we can add is the complexified FI-parameter ($t=r-i\widetilde{\theta}$) term:
\begin{equation}\label{}
  {\cal L}_{{\rm FI}}=-\frac{1}{2}\int d^{2}\Theta t\Sigma+h.c.
\end{equation}

Finally, we would like to comment that the two-forms charged matters and the three-forms super gauge field are both heavy in low energies, thus there is no gauge anomaly to worry about. Although our UV ``magnetic" matter theory only affects the topological sectors of super Yang-Mills theory; however, its UV dynamics are still very interesting, and we leave the complete investigation of this theory and its applications to future work \cite{Gu:2021GW}.

\section{Conclusions and Outlooks}
The main observation in this note is to include the field strength $F_{0\cdots d}$, apart from the scalars, to parameterize the vacuum configuration in a $1+d$ gauge theory. The field strength $F_{0\cdots d}$ is Lorentz-invariant, so even the vanishing expectation value of this field could teach us the nontrivial information of vacua. For example, when a gauge theory has a $d$-form $\mathbb{Z}_{p}$ symmetry. One can observe, by the flux, its effective theory decomposes into $p$ different universes at the intermediate scale. This is a stronger selection rule than we usually think of in the local quantum field theory. Following this setup, we reviewed the existing results in the two-dimensional quantum field theory from our perspective. Furthermore, we re-studied the dynamical decomposition in the KK-reduction of three-dimensional Chern-Simons matter theory and conjectured there is no decomposition in three dimensions. Finally, we propose a four-dimensional UV gauge theory which reduces to the generalized Yang-Mills theory in the low energy. This generalized Yang-Mills was first proposed in \cite{Seiberg:2010qd} and systematically investigated in \cite{Misumi:2019dwq}, where Tanizaki and \"{U}nsal observed the decomposition feature in the generalized Yang-Mills theory. Our proposed UV gauge theory gives a dynamical understanding. \\

So far, our discussions are limited to the theory with a Lagrangian description. However, it is known that there are many non-Lagrangian theories. Thus, it would be interesting to study how the $d$-form symmetry affects the vacuum configuration of these theories. Furthermore, we mainly focus on the continuous gauge theories in this note. However, we know many orbifolded theories can not be embedded into continuous gauge theories, so the decomposition of these orbifolded theories should be understood in a different approach. Finally, quantum field theories in more than four dimensions might also have the dynamical decomposition, which we missed in this note. We leave all these to future work.

\section*{Acknowledgement}
We would like to thank Ryan Thorngren and Yifan Wang for the debate on distinguishing between the topological current and the symmetric current in quantum field theory. It pushed the author to read the reference \cite{Gaiotto:2014kfa} and was finally convinced by the Higgs mechanism of the ``generalized global symmetry" observed in \cite{Gaiotto:2014kfa}. We also thank Eric Sharpe for the discussion and Mithat \"Unsal for his talk on the generalized Yang-Mills theory.


\begin{thebibliography}{}
\bibitem{Gaiotto:2014kfa}
D.~Gaiotto, A.~Kapustin, N.~Seiberg and B.~Willett, ``Generalized Global Symmetries,''
JHEP \textbf{02}, 172 (2015)
[arXiv:1412.5148 [hep-th]].

\bibitem{Pantev:2005zs}
T.~Pantev and E.~Sharpe, ``GLSM's for Gerbes (and other toric stacks),''
Adv. Theor. Math. Phys. \textbf{10}, no.1, 77-121 (2006)
[arXiv:hep-th/0502053 [hep-th]].

\bibitem{Pantev:2005rh}
T.~Pantev and E.~Sharpe, ``Notes on gauging noneffective group actions,''
[arXiv:hep-th/0502027 [hep-th]].

\bibitem{Pantev:2005wj}
T.~Pantev and E.~Sharpe, ``String compactifications on Calabi-Yau stacks,''
Nucl. Phys. B \textbf{733}, 233-296 (2006)
[arXiv:hep-th/0502044 [hep-th]].

\bibitem{Hellerman:2006zs}
S.~Hellerman, A.~Henriques, T.~Pantev, E.~Sharpe and M.~Ando, ``Cluster decomposition, T-duality, and gerby CFT's,''
Adv. Theor. Math. Phys. \textbf{11}, no.5, 751-818 (2007)
[arXiv:hep-th/0606034 [hep-th]].

\bibitem{Gu:2021yek}
W.~Gu, D.~Pei and M.~Zhang, ``On Phases of 3d ${\cal N}=2$ Chern-Simons-Matter Theories,''
[arXiv:2105.02247 [hep-th]].

\bibitem{Tanizaki:2019rbk}
Y.~Tanizaki and M.~\"Unsal, ``Modified instanton sum in QCD and higher-groups,''
JHEP \textbf{03}, 123 (2020)
doi:10.1007/JHEP03(2020)123
[arXiv:1912.01033 [hep-th]].

\bibitem{Komargodski:2020mxz}
Z.~Komargodski, K.~Ohmori, K.~Roumpedakis and S.~Seifnashri, ``Symmetries and strings of adjoint QCD$_{2}$,''
JHEP \textbf{03}, 103 (2021)
[arXiv:2008.07567 [hep-th]].

\bibitem{Sharpe:2014tca}
E.~Sharpe, ``Decomposition in diverse dimensions,''
Phys. Rev. D \textbf{90}, no.2, 025030 (2014)
[arXiv:1404.3986 [hep-th]].

\bibitem{Sharpe:2015mja}
E.~Sharpe, ``Notes on generalized global symmetries in QFT,''
Fortsch. Phys. \textbf{63}, 659-682 (2015)
[arXiv:1508.04770 [hep-th]].

\bibitem{Sharpe:2021srf}
E.~Sharpe, ``Topological operators, noninvertible symmetries and decomposition,''
[arXiv:2108.13423 [hep-th]].

\bibitem{Coleman:1976uz}
S.~R.~Coleman, ``More About the Massive Schwinger Model,''
Annals Phys. \textbf{101}, 239 (1976)

\bibitem{Witten:1993yc}
E.~Witten, ``Phases of N=2 theories in two-dimensions,''
Nucl. Phys. B \textbf{403}, 159-222 (1993)
[arXiv:hep-th/9301042 [hep-th]].

\bibitem{Hori:2000kt}
K.~Hori and C.~Vafa, ``Mirror symmetry,''
[arXiv:hep-th/0002222 [hep-th]].

\bibitem{Coleman:1973ci}
S.~R.~Coleman, ``There are no Goldstone bosons in two-dimensions,''
Commun. Math. Phys. \textbf{31}, 259-264 (1973)

\bibitem{Mermin:1966fe}
N.~D.~Mermin and H.~Wagner, ``Absence of ferromagnetism or antiferromagnetism in one-dimensional or two-dimensional isotropic Heisenberg models,''
Phys. Rev. Lett. \textbf{17}, 1133-1136 (1966)

\bibitem{Morrison:1994fr}
D.~R.~Morrison and M.~R.~Plesser,``Summing the instantons: Quantum cohomology and mirror symmetry in toric varieties,''
Nucl. Phys. B \textbf{440}, 279-354 (1995)
[arXiv:hep-th/9412236 [hep-th]].

\bibitem{Gu:2018fpm}
W.~Gu and E.~Sharpe, ``A proposal for nonabelian mirrors,''
[arXiv:1806.04678 [hep-th]].

\bibitem{Chen:2018wep}
Z.~Chen, W.~Gu, H.~Parsian and E.~Sharpe, ``Two-dimensional supersymmetric gauge theories with exceptional gauge groups,''
Adv. Theor. Math. Phys. \textbf{24}, no.1, 67-123 (2020)
[arXiv:1808.04070 [hep-th]].

\bibitem{Gu:2020ivl}
W.~Gu, E.~Sharpe and H.~Zou, ``Notes on two-dimensional pure supersymmetric gauge theories,''
JHEP \textbf{04}, 261 (2021)
[arXiv:2005.10845 [hep-th]].

\bibitem{Seiberg:2010qd}
N.~Seiberg, ``Modifying the Sum Over Topological Sectors and Constraints on Supergravity,''
JHEP \textbf{07}, 070 (2010)
[arXiv:1005.0002 [hep-th]].

\bibitem{Komargodski:2017dmc}
Z.~Komargodski, A.~Sharon, R.~Thorngren and X.~Zhou, ``Comments on Abelian Higgs Models and Persistent Order,''
SciPost Phys. \textbf{6}, no.1, 003 (2019)
[arXiv:1705.04786 [hep-th]].

\bibitem{Anber:2018jdf}
M.~M.~Anber and E.~Poppitz, ``Anomaly matching, (axial) Schwinger models, and high-T super Yang-Mills domain walls,''
JHEP \textbf{09}, 076 (2018)
[arXiv:1807.00093 [hep-th]].

\bibitem{Armoni:2018bga}
A.~Armoni and S.~Sugimoto, ``Vacuum structure of charge k two-dimensional QED and dynamics of an anti D-string near an O1-plane,''
JHEP \textbf{03}, 175 (2019)
[arXiv:1812.10064 [hep-th]].

\bibitem{Misumi:2019dwq}
T.~Misumi, Y.~Tanizaki and M.~\"Unsal, ``Fractional $\theta$ angle, 't Hooft anomaly, and quantum instantons in charge-$q$ multi-flavor Schwinger model,''
JHEP \textbf{07}, 018 (2019)
[arXiv:1905.05781 [hep-th]].

\bibitem{Cherman:2020cvw}
A.~Cherman and T.~Jacobson, ``Lifetimes of near eternal false vacua,''
Phys. Rev. D \textbf{103}, no.10, 105012 (2021)
[arXiv:2012.10555 [hep-th]].

\bibitem{Andreini:2008AJT}
E.~Andreini, YF,~Jiang, and HH. ~Tseng. "On Gromov-Witten theory of root gerbes." [arXiv:0812.4477 (2008)].

\bibitem{Andreini:2009AJT}
E.~Andreini, YF,~Jiang, and HH. ~Tseng. "Gromov-Witten theory of product stacks." Communications in Analysis and Geometry
Volume 24, Number 2, 223–277, 2016 [arXiv:0905.2258 (2009)].


\bibitem{Auroux:2004AKO}
D.~Auroux, L.~Katzarkov, and D. ~Orlov. “Mirror symmetry for weighted projective planes and their noncommutative deformations.” Annals of Mathematics 167 (2004): 867-943. [arXiv:math/0404281v1].


\bibitem{Govindarajan:2000ef}
S.~Govindarajan, T.~Jayaraman and T.~Sarkar, ``On D-branes from gauged linear sigma models,''
Nucl. Phys. B \textbf{593}, 155-182 (2001)
doi:10.1016/S0550-3213(00)00611-8
[arXiv:hep-th/0007075 [hep-th]].

\bibitem{Hori:2000ic}
K.~Hori, ``Linear models of supersymmetric D-branes,''
[arXiv:hep-th/0012179 [hep-th]].

\bibitem{Hellerman:2001ct}
S.~Hellerman and J.~McGreevy, ``Linear sigma model toolshed for D-brane physics,''
JHEP \textbf{10}, 002 (2001)
[arXiv:hep-th/0104100 [hep-th]].

\bibitem{Herbst:2008jq}
M.~Herbst, K.~Hori and D.~Page, ``Phases Of N=2 Theories In 1+1 Dimensions With Boundary,''
[arXiv:0803.2045 [hep-th]].

\bibitem{Hori:2013ika}
K.~Hori and M.~Romo, ``Exact Results In Two-Dimensional (2,2) Supersymmetric Gauge Theories With Boundary,''
[arXiv:1308.2438 [hep-th]].

\bibitem{Leistner:2012ika}
D. ~Halpern-Leistner. "The derived category of a GIT quotient." Journal of the American Mathematical Society 28.3 (2015): 871-912.

\bibitem{Kapustin:2002bi}
A.~Kapustin and Y.~Li, ``D branes in Landau-Ginzburg models and algebraic geometry,''
JHEP \textbf{12}, 005 (2003)
[arXiv:hep-th/0210296 [hep-th]].

\bibitem{Aganagic:2001uw}
M.~Aganagic, K.~Hori, A.~Karch and D.~Tong, ``Mirror symmetry in (2+1)-dimensions and (1+1)-dimensions,''
JHEP \textbf{07}, 022 (2001)
[arXiv:hep-th/0105075 [hep-th]].

\bibitem{Intriligator:2013lca}
K.~Intriligator and N.~Seiberg, ``Aspects of 3d N=2 Chern-Simons-Matter Theories,''
JHEP \textbf{07}, 079 (2013)
[arXiv:1305.1633 [hep-th]].

\bibitem{Witten:1999ds}
E.~Witten, ``Supersymmetric index of three-dimensional gauge theory,''
[arXiv:hep-th/9903005 [hep-th]].

\bibitem{Gu:2020zpg}
W.~Gu, L.~Mihalcea, E.~Sharpe and H.~Zou, ``Quantum K theory of symplectic Grassmannians,''
[arXiv:2008.04909 [hep-th]].

\bibitem{Kapustin:2014gua}
A.~Kapustin and N.~Seiberg, ``Coupling a QFT to a TQFT and Duality,''
JHEP \textbf{04}, 001 (2014)
[arXiv:1401.0740 [hep-th]].

\bibitem{Tachikawa:2017gyf}
Y.~Tachikawa, ``On gauging finite subgroups,''
SciPost Phys. \textbf{8}, no.1, 015 (2020)
[arXiv:1712.09542 [hep-th]].

\bibitem{WB92}
J. ~Wess and J. ~Bagger, $Supersymmetry And Supergravity$, Princeton University Press (second edition, 1992).

\bibitem{Gates:1980ay}
S.~J.~Gates, Jr., ``SUPER P FORM GAUGE SUPERFIELDS,''
Nucl. Phys. B \textbf{184}, 381-390 (1981)

\bibitem{Gu:2021GW}
In preparation.
 \end{thebibliography}
\end{document}